\documentclass[aps,pre,twocolumn,floats]{revtex4}
\usepackage{epsfig}

\begin{document}

\newcommand{\tbox}[1]{\mbox{\tiny #1}}
\newcommand{\half}{\mbox{\small $\frac{1}{2}$}}
\newcommand{\sfrac}[1]{\mbox{\small $\frac{1}{#1}$}}
\newcommand{\mbf}[1]{{\mathbf #1}}


\title{
Dephasing due to the interaction
with chaotic degrees of freedom
}

\author{Doron Cohen}

\affiliation{
Department of Physics,
Ben-Gurion University,
Beer-Sheva 84105, Israel
}

\date{March 2001, revised October 2001}

\begin{abstract}
We consider the motion of a particle,
taking into account its interaction
with environmental degrees of freedom.
The dephasing time is determined
by the nature of the environment,
and depends on the particle velocity.
Our interest is in the case where the
environment consists of few chaotic
degrees of freedom. We obtain results
for the dephasing time, and compare them
with those of the effective-bath approach.
The latter approach is based on the
conjecture that the environment can be
modelled as a collection of infinitely
many harmonic oscillators. The work is
related to studies of driven systems,
quantum irreversibility, and fidelity.
The specific model that we consider
requires the solution of the problem
of a particle-in-a-box with moving wall,
whose 1D version is related to
the Fermi acceleration problem.
\end{abstract}

\maketitle


\section{Introduction}

Determining the dephasing (decoherence) time
$\tau_{\varphi}$ for a particle $(x,p)$
that interacts with an environmental
degrees of freedom $(Q,P)$ is a central theme in
quantum physics. In the absence of such interaction
the $x$ motion is coherent, and interference should be
taken into account. This means, from semiclassical
point of view, that at least two trajectories
$x(\tau)=x_{\tbox{A}}(\tau)$ and $x(\tau)=x_{\tbox{B}}(\tau)$ have
a leading contribution to the probability to travel,
say, from $x(0)$ to $x(t)$, as in the prototype example
of the two slit experiment.

The purpose of this Paper is to discuss the
case where the particle interacts with few
chaotic degrees of freedom. To be more specific
we consider a box/piston model which is defined
in Section~2 below.
It is already known that in the classical descriptions
such interaction leads to dissipation,
and the motion of the particle is described
by the standard Langevin equation \cite{jar}.
Our aim is to explore the quantum-mechanical consequences
of this interaction, and in particular to determine
the dephasing time.

A related aim of this paper is to test
the effective-bath conjecture, namely,
that any type of environment can be modelled
as a large collection of harmonic oscillators.
This {\em conjecture} is implied
by leading order perturbation theory \cite{FV}.
We shall explain how this conjecture
can be applied, in practice, in order
to re-derive our results for the dephasing time.

Some readers may wonder whether dephasing is not
just the entanglement of the particle with some
other degrees of freedom.
In such case even a one-spin environment
can provide dephasing (decoherence).
However, we would like to adopt a more
restrictive definition of dephasing,
that involves the notion of irreversibility (see Appendix).

At first sight it seems that for having irreversibility
one needs "infinity". This point of view is
emphasized in Ref.\cite{buttiker}:
Irreversibility can be achieved by having the
infinity of the bath (infinitely many oscillators),
or of space (a lead that extends up to infinity).
The present Paper is based on the observation that
also chaos provides irreversibility.
We do not need "infinity" in order to have
"irreversibility". This conceptual point is further
clarified in the Appendix.

The present work is strongly related to so-called
studies of quantum irreversibility \cite{peres}
and fidelity \cite{jalabert},
and hence to studies of wavepacket dynamics \cite{heller},
decay of the survival amplitude, and the
parametric theory of the local density
of states \cite{wls,lds,prm}.
We explain this point in Section~7.
We would like to emphasize that all these
studies share a common conceptual framework.
The common idea \cite{crs} is that we have to
distinguish between perturbative and non-perturbative
regimes, and that the semiclassical (sub) regime
is contained within the non-perturbative regime.

The study of the box/piston model reduces to the
analysis of the problem of a driven particle.
Namely, the problem of a particle-in-a-box with moving wall.
This is possibly the simplest example for illustrating the
idea of having different dynamical regimes.
The velocity of the wall is $V$. We shall explain that
in the one-dimensional (1D) box problem
(also known as the problem of infinite well
with moving wall \cite{infwell_a,infwell_b}),
there are two quantum-mechanical $V$ regimes:
An adiabatic $V$ regime, and a non-perturbative
semiclassical $V$ regime. In the general box/piston problem
there are three regime: The additional $V$ regime is
a perturbative regime, where we can apply Fermi golden
rule in order to describe the stochastic energy spreading.

\begin{figure}
\centerline{\epsfig{figure=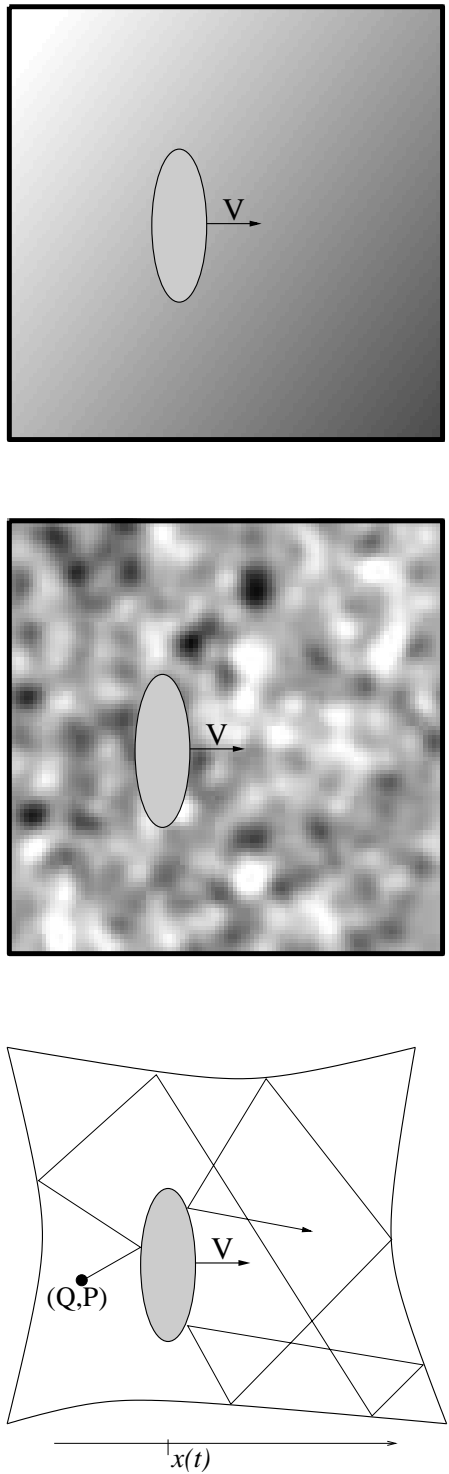,width=6cm}}
\vspace{0.5cm}
\caption{
(a) The Brownian particle in the ZCL model
experiences a fluctuating homogeneous field of force.
(b) In case of the DLD model the fluctuating field
is farther characterized by a finite correlation distance.
(c) The Brownian motion is induced due to the
interaction with chaotic degrees of
freedom. $x(t)$ is the (classical) position of the
Brownian particle.
}
\end{figure}

\section{Interaction with chaos}

One can imagine, in principle, a "zoo" of models that
describe interaction of a particle with the environment.
However, following Caldeira and Leggett \cite{zcl},
the guiding philosophy is to construct "ohmic models"
that give Brownian motion (described by the standard
Langevin equation) in the classical limit.
Three families of models are of particular interest:
\begin{itemize}
\item Interaction with chaos.
\item Interaction with harmonic bath.
\item Interaction with random-matrix bath.
\end{itemize}
The second type of modelling leads to the introduction
of the Diffusion-Localization-Dissipation (DLD) model \cite{dld}.
The familiar Zwanzig-Caldeira-Leggett (ZCL) model \cite{zcl}
can be regarded as a special limit of the DLD model.
The physics of the ZCL and of the DLD model is illustrated
in Fig.1a and Fig.1b respectively. The ZCL model describes a motion
under the influence of a fluctuation homogeneous filed
of force. In case of the DLD model the fluctuation field
is further characterized by a finite correlation distance.
We shall come back to these models in Section~10.
For completeness we note that random-matrix modelling of the
environment, in the regime where it has been solved \cite{rmt},
leads to the same results as those obtained for the DLD model.

In this paper we are going to consider the
case where the particle ($x$) interacts with
few chaotic degrees of freedom. It is well
known that {\em classically} such an interaction
has the same effect as that of coupling to
an ohmic bath \cite{jar}. Quantum mechanically much
less is known \cite{vrn}.

We shall analyze the prototype toy model
which is illustrated in Fig.1c. The dynamical
variable $x$ represents the position of a large
(`Brownian') particle. The motion of this particle
is affected by collisions with a small (`gas') particle.
Thus $Q$ is the position of the gas particle and
$P$ is the conjugate momentum. The motion of the
gas particle is assumed to be chaotic and its collisions
with the Brownian particle are assumed to be elastic.
The typical time between successive collisions
will be denoted by $\tau_0 = L/v_{\tbox{E}}$,
where $v_{\tbox{E}}$ is the typical velocity
of the gas particle.
The typical kinetic energy of the gas particle
$E=\half m v_{\tbox{E}}^2$ has the same significance
as the temperature $T$ in the ZCL/DLD models.
(For sake of exact comparison we should assume that
$E$ has a canonical distribution, but for the purpose
of this presentation we prefer to assume that it has
some well-defined microcanonical value).

\section{Dynamical regimes}

The results of the forthcoming analysis depend on the
typical velocity $V=|\dot{x}|$ of the Brownian particle.
A fixed assumption of this Paper is that $V$ is slow
in a classical sense, meaning $V \ll v_{\tbox{E}}$.
In the quantum mechanical analysis we are going to
distinguish the following quantal $V$ regimes:
\begin{eqnarray} \label{e3}
\mbox{adiabatic regime: \ \ \ \ }
V \ll \left(\frac{\lambda_{\tbox{E}}^{d{-}1}}
{A}\right)^{\frac{3}{2}}
\frac{\hbar}{mL}
\\ \label{e1}
\mbox{non-perturbative regime: \ \ \ \ \ \ \ \ }
V \gg \frac{\hbar}{mL}
\end{eqnarray}
where $d$ is the dimensionality of the box,
and $A$ is the effective area of the $d{-}1$
dimensional surface of the Brownian particle.
The De-Broglie wavelength $\lambda_{\tbox{E}}$
of the gas particle is defined as in Eq.(\ref{e7}).
For $d>1$ the above two $V$ regimes are well separated,
and we have a third `perturbative' regime where
$V$ is small ($\ll \hbar/(mL)$) but non-adiabatic.

The naive semiclassical expectation,
regarding the system of Fig.1c, is to have
a loss of coherence upon collision. In other
words, we expect to have
\begin{eqnarray} \label{e0}
\tau_{\varphi} \ \ = \ \ \tau_0
\end{eqnarray}
We want to go beyond this naive expectation;
to obtain specific results for the dephasing time;
and to compare them with the prediction which is based
on the effective-bath conjecture.

Specifically, we are going to claim that
the naive semiclassical result is valid
only in the non-perturbative regime.
Otherwise, in the perturbative regime,
we get that the dephasing time is much larger:
\begin{eqnarray} \label{e2}
\tau_{\varphi} \ \ \approx \ \
\left(\frac{L\lambda_{\tbox{E}}^2}
{v_{\tbox{E}}V^2}\right)^\frac{1}{3}
\end{eqnarray}

Finally, in the adiabatic regime,
to the degree that adiabaticity is maintained,
there is no dephasing at all. However, in practice
one should take mainly Landau-Zener transitions
into account \cite{wilk}
(see also Section 20 of \cite{frc}).
This leads to a finite dephasing time
$\tau_{\varphi}  \propto (1/V)^{1+(\beta/2)}$,
where the value of $\beta$ depends on the level
spacing statistics.


\section{Definition of the dephasing factor}

The Hamiltonian of the system$+$environment
can be written as
\begin{eqnarray} \label{e_10}
{\cal H}_{\tbox{total}}(x,p,Q,P) =
{\cal H}_0(x,p) + {\cal H}(Q,P;x)
\end{eqnarray}
Using this Hamiltonian we should be able
to calculate, in principle, any transport
property. To test whether "interference"
is present we should be able to control
this interference. For example in a two slit
geometry we control the relative position of
the detector, while in Aharonov-Bohm ring
geometry (see Appendix) we control the
magnetic flux via the ring.

A more restricted definition of dephasing
can be obtained within the framework of the
Feynman-Vernon formalism \cite{FV}. After elimination
of the bath degrees of freedom, one
ends up with a {\em double} path-integral for the
transport. This expression is not very
illuminating unless the $x$ motion can be treated
semiclassically \cite{dph}. In such case
it becomes a double sum over "classical"
trajectories, and we can interpret the
"off diagonal" terms as responsible for the
interference effect. Due to the elimination
of the bath degrees of freedom, each interference
term is multiplied by a factor which is
known as the "influence functional".
The absolute value of the influence functional is
defined as the "dephasing factor".

The influence functional is traditionally
calculated for linear coupling to harmonics
oscillators, assuming that they are initially
prepared in a canonical thermal state.
However, we would like to consider the case of
interaction with chaotic degrees of freedom,
and we would like to assume that the "environment"
is initially prepared in a pure state $\Psi_0$.
Thus the "bath" is characterize by its
microcanonical energy $E$ rather by its
temperature $T$. [Obviously one can obtain the
thermal case by canonical averaging over $E$.]

In order to calculate the influence functional
one considers the evolution which is generated
by the time dependent Hamiltonian
\begin{eqnarray} \label{e_11}
{\cal H} \ \ = \ \ {\cal H}(Q,P;x(\tau)),
\end{eqnarray}
for the particular (interfering) trajectories
$x(\tau)=x_{\tbox{A}}(\tau)$ and $x(\tau)=x_{\tbox{B}}(\tau)$
that connects $x(0)$ and $x(t)$.
The initial preparation of the environment
is represented by a wavefunction $\Psi_0$,
while the final state is either $\Psi(t)=\Psi_{\tbox{A}}$
or $\Psi(t)=\Psi_{\tbox{B}}$.
The overlap of the two possibilities
is known as the influence functional
\begin{eqnarray} \label{e_12}
c(t) \ \ = \ \ F[x_{\tbox{A}},x_{\tbox{B}}] \ \ = \ \
\langle \Psi_{\tbox{B}} | \Psi_{\tbox{A}} \rangle
\end{eqnarray}
The dephasing factor is defined as the absolute value
of the influence functional.

We have introduced above the alternate notation $c(t)$
in order to emphasizes the time dependence of the wavefunction
overlap. In order to make $c(t)$ a well-defined quantity,
one should pre-define the statistical properties of the
interfering trajectories as a function of $t$.
See \cite{qbm} for mathematical details. For the
purpose of the present Paper one can assume that
typical interfering trajectories are ballistic,
and characterized by their velocity $V$, and by their
(maximal) separation $A \sim Vt$. This motion scheme
provides "maximal dephasing". For other schemes
of motion the maximal separation scales differently.
For example for two-slit geometry $A$ is the separation
between the slits. For diffusive trajectories
$A \propto \sqrt{t}$. Note however that a full statistical
specification is needed in each case \cite{qbm}.

\section{Remark: Interpretation of the dephasing factor}

From the definition of the influence functional it is clear
that it reflects the probability to "leave a trace" in the
environment. In case of the DLD model (see Section~10)
this "trace" can be further interpreted as leaving an excitation
along the way. For critical discussion of this point
see Appendix C of \cite{qbm}. In the more general case
the notion of "leaving a trace" is somewhat ambiguous.
All we can say is that decoherence means that the environment
is left in different (orthogonal) states depending on the
trajectory that is taken by the particle.

The law of "action and reaction" holds also in the world
of decoherence studies. Feynman and Vernon have realized that
the dephasing factor can be re-interpreted as representing
the effect of a c-number noise source. From this point
of view the decoherence is due to the "scrambling" of the
relative phase by this noise. Hence the reason for using
the term "dephasing" as a synonym for "decoherence".

The advantage of the latter point of view is that it can
give further insight regarding the physics of our results.
Namely, we shall see in Section~11
that Eq.(\ref{e0}) and Eq.(\ref{e2}) have
an effective DLD-model, and effective ZCL-model interpretations
respectively. From the particle's dynamics point of view
this corresponds to "scattering mechanism"
and to "spreading mechanism" as discussed in \cite{qbm}.

\begin{figure}
\centerline{\epsfig{figure=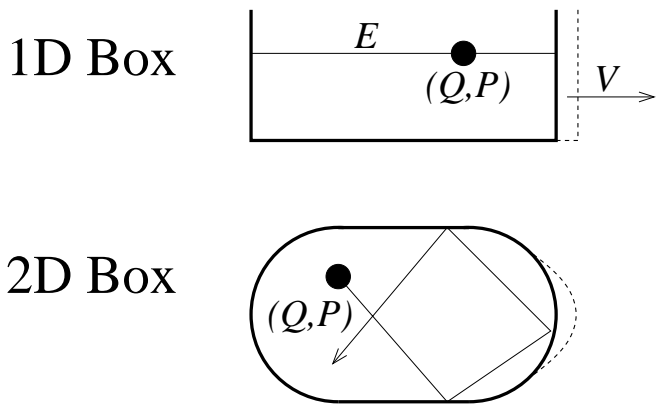,width=\hsize}}
\vspace{0.8cm}
\caption{
A `gas' particle in a box is driven by moving
the wall. This is essentially the same as
Fig.1c, but now the moving wall is {\em not}
regarded as a dynamical entity.
(a) illustrates the case of one-dimensional
box, while (b) is for {\em chaotic} box.
(Hence, in the latter case, the box should be
at least two-dimensional).
}
\end{figure}

\section{Determination of the dephasing time}

Within the semiclassical framework,
the problem of dephasing reduces to the more
restricted problem of studying the dynamics of a
time dependent Hamiltonian (Eq.(\ref{e_11})).

By definition, in order to have coherence ($|c(t)|\sim 1$),
the wavefunction $\Psi(t)$ should contain a component
which is independent of the particular way
in which $x(\tau)$ evolves from the initial $x(0)$
to the final $x(t)$. Loss of coherence means
$|c(t)| \ll 1$, which can be written as
$t < \tau_{\varphi}$. This constitutes a practical
definition of the dephasing time~$\tau_{\varphi}$.

It should be clear that the toy system of Fig.1c
is completely equivalent to the toy system which
is illustrated in Fig.2. Namely, what we have to
analyze is the evolution which is generated by
${\cal H}(Q,P;x(\tau))$, where $x$ controls the deformation
of the boundary. In particular we should determine
whether $\Psi(t)$ possesses a trajectory-independent
component that is determined only by the endpoints
$x(0)$ and $x(t)$.

The rest of this Paper is organized as follows:
In Section~8 we illuminate the key ingredients in the
analysis by considering a one-dimensional example;
In Section~9 we outline the derivation
of Eq.(\ref{e0}) and Eq.(\ref{e2})
using the core-tail picture which has been developed
in \cite{frc}, and which is supported by our recent numerical
studies \cite{prm,lds,wlm};
In Section~11  we compare the results
to those of the effective-bath approach.

\section{Digression: Relation to fidelity and LDOS studies}

The definition of $c(t)$ can be re-written formally as
\begin{eqnarray} \label{e_13}
c(t) = \left \langle \Psi_0 | \ U[x_{\tbox{B}}(\tau)]^{-1}
U[x_{\tbox{A}}(\tau)] \ | \Psi_0 \right\rangle
\end{eqnarray}
where $U[x(\tau)]$ is the evolution operator
due to a driving "pulse" $x(\tau)$.
Thus $c(t)$ can be re-interpreted as the probability
amplitude to come back to the initial state at the end
of a "driving cycle". This quantity, which quantifies
the "fidelity" of the driving cycle, has been suggested
by Ref.\cite{peres} to be a measure for quantum irreversibility.

To be more precise, the original definition of the
fidelity in Ref.\cite{peres} assumes rectangular "pulses".
This means, without loss of generality,
that $x_{\tbox{A}}(\tau)=0$ and
$x_{\tbox{B}}(0<\tau<t)=A=\mbox{const}$.
Thus the time variation of $c(t)$ depends on the
amplitude $A$ of the "perturbation". Further simplification
is obtained if $\Psi_0$ is assumed to be an eigenstate
of the unperturbed Hamiltonian ${\cal H}_0={\cal H}(Q,P;0)$.
In such case $c(t)$ is known as the survival amplitude,
and we get
\begin{eqnarray} \label{e_14}
|c(t)|^2 = \left|\left\langle \Psi_0 \Big|
\exp\left( -\frac{i}{\hbar}{\cal H}t \right)
\Big| \Psi_0 \right\rangle\right|^2
\end{eqnarray}
where ${\cal H}={\cal H}(Q,P;A)$ is the perturbed
Hamiltonian. The study of the survival probability
is also known as "wavepacket dynamics". Note that
the survival probability is the Fourier transform
of the of the local density of states (LDOS).
Hence the study of "wavepacket dynamics" can be
reduced to LDOS study. For an introduction to this
subject see \cite{heller}.

The fidelity in general, and the survival amplitude
in particular have similar physics.
In Ref.\cite{peres} it is explained that
if $A$ is small (in the sense of standard
first order perturbation theory), then the decay
of $c(t)$ has a Gaussian time dependence
due to a statistical effect. For larger $A$ we
get exponential time dependence as in Wigner theory.
On the other hand in semiclassical circumstances \cite{heller,wls}
the decay may become {\em perturbation independent}.
This idea was generalized in \cite{jalabert}.
In such case the rate of the decay is determined by the
stability of the classical motion, and is characterized
by the Lyapunov exponent.
A unified picture of the crossover from the
perturbative $A$ regime to the semiclassical $A$ regime
has been presented in \cite{wls}
and has been generalized in \cite{philip}.

Does the study of "rectangular pulses"
constitute a good bridge for developing a
general theory for fidelity?
The answer is definitely not.
An essential ingredient in the theory of
driven system is the rate $V$ in which the
parameter $x$ is being changed in time.
Thus, rather than talking about $A$ regimes,
we should talk about $V$ regimes, as in the
present Paper. The general theory becomes
more complicated \cite{crs,frc,vrn},
but the physical picture is similar in spirit.
The box/piston model that we are going
to analyze in the next sections is possibly
the simplest demonstration for the applicability
of the ideas which were presented in \cite{crs,frc,vrn}.
More generally we should talk about $(V,A)$ regimes,
as in the theory of periodically driven mesoscopic
systems \cite{rsp}. The latter issue is beyond
the scope of the present Paper.

\section{Analysis of the 1D box problem}

Consider the one-dimensional system of Fig.2a.
The classical analysis of the dynamics is trivial.
Each time that the gas particle collides with the
moving wall it loses energy: Upon collision its
velocity undergoes a change $v \mapsto -v+2V$
and therefore the change in energy is $dE_{\tbox{col}}=-2mvV$.

Now we want to analyze the dynamics quantum-mechanically.
This turns out to be less trivial. Past studies of this
model \cite{infwell_a,infwell_b}, which are related to the interest
in the Fermi accelerator problem \cite{jose}, were focused
on the issue of finding stationary solutions.
To the best of our knowledge, the time dependent picture has
not been explored.

Let $|n(x)\rangle$ denote the eigenstates of the box Hamiltonian
${\cal H}(Q,P;x)$. The expansion of the wavefunction in
this $x$-dependent basis is $\Psi(t)=\sum_n a_n(t)|n(x(t))\rangle$.
The expansion coefficients $a_n(t)$ are the
probability amplitudes to find the particle in the energy
level $n$ after time $t$.  One easily obtains the equation:
\begin{eqnarray} \label{ee5}
\frac{da_n}{dt} =
- \frac{i}{\hbar} E_n a_n
- \frac{V}{L} \sum_{m (\ne n)} \frac{2nm}{n^2-m^2} \ a_m
\end{eqnarray}
Let us assume that the initial preparation
is $a_n(0)=\delta_{nm}$.
The mean level spacing for the 1D box is
$\Delta = \pi\hbar v_{\tbox{E}}/L$.
If $dE_{\tbox{col}} \ll \Delta$ one finds out,
by inspection of Eq.(\ref{ee5}),
that the dynamics is adiabatic,
meaning that $a_n(t)\sim\delta_{nm}$.
On the other hand, if $dE_{\tbox{col}} \gg \Delta$,
one expects to find a semiclassical transition
$E \mapsto E + dE_{\tbox{col}}$.

How can we explain the $E \mapsto E + dE_{\tbox{col}}$
transition from quantum-mechanical point of view?
For this purpose we can adopt the
core-tail picture of Ref.\cite{frc}.
The core-tail picture is a generalization
of Fermi-golden-rule picture:
The `core' consists of the levels that are
mixed non-perturbatively; The `tail' is formed by
first order transitions from the core.

The analysis is carried out in two steps.
The first step is to analyze the {\em parametric evolution}
which is associated with Eq.(\ref{ee5}).
This means to solve Eq.(\ref{ee5}) without
the first term in the RHS. (This is the so-called sudden limit).
Obviously the resultant $\tilde{a}_n(t)$
is a function of $\delta x =Vt$, while
$V$ by itself makes no difference. The solution
depends only  on the endpoints $x(0)$ and $x(t)$.
The second step is to analyze
the actual time evolution. This means to take into account
the effect of the first term in the RHS  of Eq.(\ref{ee5}),
and to understand how the resultant $a_n(t)$ differs
from $\tilde{a}_n(t)$.

By careful inspection of Eq.(\ref{ee5}) one observes
that a level is mixed with the next level whenever
the wall is displaced an additional distance $\lambda_{\tbox{E}}/2$.
This effect can be regarded as `parametric'.
Further inspection reveals that this non-perturbative
(parametric) effect modulates the core-to-tail transition
amplitude (see remark \cite{rmrk1}).
The modulation frequency is
$\omega = 2\times 2\pi/(\lambda_{\tbox{E}}/V)$.
This frequency drives a core-to-tail
resonance transition  $|dE| = \hbar\omega$,
in agreement with the semiclassical expectation.

\section{Analysis of the general box problem}

The strength of the core-tail picture is that
it can be used to analyze the more general case
which is illustrated in Fig.2b. We assume that
the motion of the particle inside the box is
chaotic in the classical limit. The quantum-mechanical
analysis follows the same steps as in the 1D problem,
and requires the use of results that we have
obtained in previous publications (mainly \cite{prm}).
In order to keep the presentation illuminating,
and trying to avoid duplications,  we shall just
sketch the derivation.

On the basis of the analysis of Ref.\cite{prm}
we recall that there are two relevant parametric
scales:
\begin{eqnarray} \label{e_6}
\delta x_c^{\tbox{qm}} \ \ &\approx& \ \
\left(\frac{\lambda_{\tbox{E}}^{d{-}1}}{A}
\right)^{\half}\times \lambda_{\tbox{E}}
\\ \label{e7}
\delta x_c^{\tbox{cl}} \ \ &=& \ \
\lambda_{\tbox{E}} \ \ = \ \
(2\pi\hbar)/(mv_{\tbox{E}})
\end{eqnarray}
The first parametric scale tells us what is the
displacement $\delta x$ which is required to mix
neighboring levels. The second parametric scale
determines what is the 'linear' regime
of this deformation process, and marks
the parametric crossover from the perturbative
to the (non-universal) semiclassical regime.
For more details see \cite{prm}.

The existence of two distinct parametric scales
implies \cite{frc} that there are three $V$ regimes in the problem:
The most trivial one is the adiabatic regime (Eq.(\ref{e3})),
which is defined via the requirement
\begin{eqnarray}
Vt_{\tbox{H}} \ \ \ll \ \ \delta x_c^{\tbox{qm}}
\end{eqnarray}
where $t_{\tbox{H}}=2\pi\hbar/\Delta$ is the Heisenberg time
for recurrences. From this definition it is clear that
quantum recurrences start before the probability
goes to other levels. As a results, in leading order
description, the probability remains concentrated all
the time in the initial level.
From now on we assume without saying
that we are in the non-adiabatic $V$ regime(s).

The non-perturbative semiclassical
regime (Eq.(\ref{e1})) is defined by the requirement
\begin{eqnarray}
V\tau_0 \ \ \gg \ \ \delta x_c^{\tbox{cl}}
\end{eqnarray}
If $V$ is non-adiabatic on the one hand, and not large
in the sense of Eq.(\ref{e1}) on the other hand, then,
using the terminology of \cite{frc},
we are in the (extended) perturbative regime.
In the 1D case $\delta x_c^{\tbox{qm}}$
and $\delta x_c^{\tbox{cl}}$ coincide, and therefore
the perturbative regime is absent!
We turn now to explain the mechanism for
energy spreading in the (extended) perturbative
regime. We shall call it the "Fermi golden rule"
picture.  After that we explain how to
obtain the semiclassical picture that arise in
the non-perturbative regime.

After displacement $\delta x = Vt$ of the wall,
the number of levels that become mixed
non-perturbatively \cite{frc} is
$(\delta x / \delta x_c^{\tbox{qm}} )^2$.
[In general \cite{lds} there may be non-universal
deviations from this quadratic growth,
leading straightforwardly to possible
generalization of our results].
Thus we can define the core width (in energy units) as
\begin{eqnarray} \label{e8}
\Gamma(\delta x) \ \ = \ \
\left(\frac{\delta x}
{\delta x_c^{\tbox{qm}}}\right)^2
\times \Delta \end{eqnarray}
where $\Delta$ is the mean level spacing.
As explained in Ref.\cite{prm}, this non-perturbative
mixing proceeds as long as $\delta x < \delta x_c^{\tbox{cl}}$
and provided $\Gamma \ll \hbar/t$.

In the (extended) perturbative regime
the inequality $\Gamma \ll \hbar/t$
breaks down before $\delta x \sim \delta x_c^{\tbox{cl}}$.
This determines the dephasing time of Eq.(\ref{e2}).
Coherence is maintained for $t<\tau_{\varphi}$
because most of the probability is still concentrated
in the core, whose evolution is of `parametric' nature.
In other words: as long as the core is not
resolves ($\Gamma \ll \hbar/t$), its evolution
depends only on the endpoints $x(0)$ and $x(t)$.

What about the evolution of the tails?
A general argumentation (see Sec.16 of Ref.\cite{frc})
implies that as long as $\delta x < \delta x_c^{\tbox{cl}}$
the core-to-tail transitions are not affected by the
local non-perturbative mixing of the levels
(no modulation of the core-to-tail transition
amplitude). However, in the (extended) perturbative
regime the core is resolved much before
we get to $\delta x \sim \delta x_c^{\tbox{cl}}$.
Consequently, we can use the Fermi-golden-rule
picture in order to describe the crossover to
stochastic diffusion in energy \cite{crs,frc,vrn}.

A different picture arises in the non-perturbative
regime. From the definition of this regime it follows
that we still have $\Gamma \ll \hbar/t$ at the time when
$\delta x \sim \delta x_c^{\tbox{cl}}$.
In this case the core saturates to a semiclassically
determined profile (see details in Sec.10 of \cite{prm}),
having the width
\begin{eqnarray}
\Gamma \ \ \sim \ \ \hbar/\tau_0
\end{eqnarray}
The time to resolve this width is $\tau_0$.
Consequently the dephasing time is simply
$\tau_{\varphi} = \tau_0$,
which is the naive semiclassical result.
As for the core-to-tail transitions: These
are  modulated as in the analysis of the
one-dimensional case. Consequently the long time
dynamics in the non-perturbative regime is
of semiclassical (rather then of Fermi-golden-rule) nature.


\section{The DLD and the ZCL models}

Following Feynman and Vernon it is common to model the
environment as a huge collection $Q=\{Q_{\alpha}\}$ of harmonic
oscillators. The advantage of such modelling  is obviously
the ability to make an exact elimination of the
environmental degrees of freedom, and to end up with
a simple path integral expression for the (reduced)
propagator of the particle.

In case of the ZCL model the interaction of the particle
($x$) with the environmental degrees of freedom
($Q_{\alpha}$) is expressed as ${\cal H}_{\tbox{int}}
= x \sum_{\alpha} c_{\alpha} Q{\alpha}$, where
$c_{\alpha}$ are coupling constants. Thus, in case
of the ZCL model the particle experiences fluctuations
of homogeneous field of force (Fig.1a).

In case of the DLD model the interaction with
$Q_{\alpha}$ is expressed as ${\cal H}_{\tbox{int}}
= \sum_{\alpha} c_{\alpha} Q_{\alpha} u(x-x_{\alpha})$,
where $u(r)$ is a short-range interaction, and
$x_{\alpha}$ is the location of the ${\alpha}$
oscillator. Thus, in case of the DLD model the
particle experiences fluctuations of disordered
field of force (Fig.1b).
A specific example for the application of the DLD model is
the dephasing of an electron due to its Coulomb interaction
with the rest of the Fermi sea \cite{imry,dph}.

The long-time classical motion of the Brownian
particle, for all three models of Fig.1,
is described by the same Langevin equation
$m\ddot{x} = -\eta\dot{x} + {\cal F}$,
where $\eta$ is called the friction coefficient,
and ${\cal F}$ can be regarded as arising from
stochastic-like fluctuating field of force.
The fluctuating force is characterized
by an intensity $\nu$ which is related to $\eta$
via a fluctuation-dissipation (FD) relation \cite{rmrk2}.
In case of the DLD model, the
fluctuating field (Fig.1b)
is further characterized by a correlation
distance $\ell$, which is determined by the range
of the interaction $u(r)$.

\section{The effective bath conjecture}

If we make the conjecture that the system
of Fig.1c is effectively described by the
DLD model, then we should substitute
\begin{eqnarray} \label{e5}
(\nu)_{\tbox{effective}} \ \ &=& \ \
m^2 v_{\tbox{E}}^3 / L
\\ \label{e6}
(\ell)_{\tbox{effective}} \ \ &=& \ \
\delta x_c^{\tbox{cl}} \ \ = \ \
\lambda_{\tbox{E}}
\end{eqnarray}
For getting Eq.(\ref{e5}) see details in Sec.7
of Ref.\cite{frc}. For getting Eq.(\ref{e6}) see
detailed discussion in Sec.11 of \cite{prm}.

The dephasing time in the high temperature
limit of the DLD model is given by \cite{dld}
\begin{eqnarray} \label{e4}
\tau_{\varphi} \ \ = \ \
\frac{\hbar^2}{\nu \ \ell^2}
\ \ \ \ \mbox{for} \ \
T \gg \hbar \frac{V}{L}
\end{eqnarray}
Upon substitution of (\ref{e5}) and (\ref{e6})
into (\ref{e4}), we get $\tau_{\varphi} = \tau_0$
which is the naive semiclassical result.
With the identification $E \leftrightarrow T$,
the high temperature condition of Eq.(\ref{e4})
becomes $V \ll v_{\tbox{E}}$, which is just the
classical slowness condition which we assume
in any case.

Eq.(\ref{e4}) becomes non-applicable
if $V\tau_{\varphi} < \ell$.
In such case the dephasing happens before
$\delta x$ gets to $\ell$ and therefore
the DLD model reduces to the ZCL model.
One easily verifies that the above distinction
between ZCL and non-ZCL regimes formally
coincides with Eq.(\ref{e1}): What appears
to be non-perturbative in case of Fig.1c
appears as a non-ZCL feature in the effective
bath description.

Is it possible to give an effective-bath
interpretation to Eq.(\ref{e2})? The answer
is positive. Using the same procedure
to estimate $\tau_{\varphi}$ for the ZCL
model as in \cite{qbm}, and upon using the
estimate $|x_{\tbox{A}}(\tau)-x_{\tbox{B}}(\tau)| \sim Vt$,
one obtains
\begin{eqnarray} \label{e_4}
\tau_{\varphi} \ \ = \ \
\left(
\frac{\hbar^2}{\nu \ V^2}
\right)^{\frac{1}{3}}
\end{eqnarray}
Substitution of (\ref{e5}) gives Eq.(\ref{e2}).
It is important to realize that these results
are `worst case estimate'. We have assumed
that only the core component is capable of maintaining
coherence.

Using the effective bath approach it is easier
to get a heuristic (phase-space based) understanding
of why $\tau_{\varphi}$ can be much longer compared with
the above estimate. In the ZCL regime, the
actual value of the dephasing time is quite
sensitive to the geometry of the interfering
trajectories. See further discussion of
dephasing via the `spreading mechanism' in \cite{qbm}.

\appendix

\section{Irreversibility versus recurrences}

In order to avoid confusion it is better to distinguish
between the notions of "irreversibility" and "recurrences".
In the first part of this Appendix we define and discuss the
issue of irreversibility, while in its second part we
define and discuss the issue of recurrences.

We say that a process is reversible if it is possible
to "undo" it. For example: consider a gas inside a
cylinder with a piston. Let us shift the piston inside.
Due to the compression the gas is heated up.
Can we undo the "heating" simply by shifting the piston
outside, back to its original position?
If the answer is yes, as in the case
of strictly adiabatic process, then we say that
the process is reversible.

Consider the prototype example of interference
in Aharonov-Bohm ring geometry. The particle can
go from the input lead to the output lead
by travelling via either arms of the ring.
This leads to interference, which can be tested
by measuring the dependence of the transmission
on the magnetic flux via the ring. Consider now
the situation where there is a spin degree of freedom
in one arm \cite{imry}. The particle can cause a spin flip
if it travels via this arm. In such case interference
is lost completely. However, this entanglement
process is completely reversible. We can undo
the entanglement simply by letting the particle
interact with the spin twice the time.
Therefore, according to our restrictive definition,
this is not a real dephasing process.

Consider now the situation where a particle
gets entangled with bath degrees of freedom.
If the bath is infinite, then the entanglement
process is irreversible, and therefore it
constitutes, according to our definition,
a dephasing process. In this paper we have
analyzed a more tricky situation where a particle
gets entangled with chaotic degrees of freedom.
The environment is finite, but due to its
chaotic nature we have irreversibility.
Hence we can talk about dephasing process.

Consider an ice-cube inside a cup of tea.
After some time it melts and disappears. But if we
wait long enough we have some probability to see
the ice-cube re-emerging due to recurrences.
The issue of recurrences becomes relevant
whenever we consider a closed (un-driven) system.
In other words, whenever we do not try to control
its evolution from the outside.

There are recurrences both in classical and quantal
physics. In the latter case the tendency for recurrences
is stronger due to the quasi-periodic nature of the
dynamics. However, if the time scale for recurrences is
long enough with respect to other relevant time scales,
then we can practically ignore these recurrences.
Actually it is useful to regard these recurrences
as "fluctuations", and to take the standpoint that
our interest is only in some "average" behavior.

If the bath is infinite, then also the time
for recurrences of the particle-bath system
becomes infinite.
On the other hand, if the bath is finite,
then we have to consider the recurrences of the
particle-bath system. These recurrences
can lead back to an un-entangled state.

In practice the time to get un-entangled
by recurrences is extremely large.
Assuming a chaotic environment, and ignoring
issues of level statistics, the time scale
for recurrences is at least the Heisenberg time
(inverse of the mean level spacing)
of the combined particle-environment system.
It scales like $\hbar^{-(d+d_0)}$ where $d_0$
is the number of degrees of freedom of the particle.

It goes without saying that the above issue
of recurrences becomes irrelevant if the
$x$ motion is treated classically.
There is however a twist to this
latter statement in the case where
the time variation of $x$ is strictly periodic.
This is due to dynamical localization effect \cite{qkr}.
Note however that dynamical localization
is a very fragile effect. Even in case that
it is found, it turns out that it manifests
itself only after a time that scales
like $\hbar^{-(1+2d)}$, which is much larger
then the Heisenberg time of the environment \cite{rsp}.


\acknowledgments{
I thank Shmuel Fishman (Technion) and Tsampikos Kottos
(MPI Gottingen) for useful discussions.
The Max Planck Institute for Physics of Complex Systems
is acknowledged for generous hospitality during the workshop
"Coherent Evolution in Noisy Environments" (Dresden, 2001),
where the present study was first presented.
}





\begin{thebibliography}{99}
\bibitem{rmrk1} For each $\lambda_{\tbox{E}}/2$ displacement,
the driving term (second term in RHS of Eq.(\ref{ee5})) changes sign.
This is due to mixing of the $m$-th level with the next level.
Within a time period $\lambda_{\tbox{E}}/V$, the driving term changes
its sign twice. The associated modulation frequency $\omega$ includes
an extra $(2{\times})$ factor reflecting that mixing happens at both
ends of the transition.
\bibitem{rmrk2}
For a canonical bath the FD relation is $\nu = 2\eta T$.
For the system of Fig.1c, assuming microcanonical preparation,
a more general version of the FD relation \cite{frc} leads
to the same expression with $T$ replaced by $(4/(d{+}1))E$.
\end{thebibliography}
\end{document}